\documentclass[12pt]{article}
\usepackage{amsmath}
\usepackage{amssymb}
\usepackage{epsfig}

\begin{document}
\newcommand{\El}{\ensuremath{\mathcal L_2(\mathbb{R})}}
\newcommand{\Av}{\ensuremath{\langle A \rangle}}
\newcommand{\Tr}{\ensuremath{\mathrm {Tr}}}
\newcommand{\Avos}{\ensuremath{\Av_{osc}}}
\newcommand{\Rho}{\ensuremath{\rho_0^{osc}}}
\newcommand{\im}{\ensuremath{\Im \mathrm{m}}}
\newcommand{\aka}{\ensuremath{a^\dag a}}
\newcommand{\eg}{\ensuremath{e^{-\frac{g^2}2}}}
\newcommand{\z}{\ensuremath{\frac 1Z}}
\newcommand{\ehp}{\ensuremath{e^{iH_0t}}}
\newcommand{\ehm}{\ensuremath{e^{-iH_0t}}}
\newcommand{\egbk}[1]{\ensuremath{e^{ig\,b^\dag \,#1}}}
\newcommand{\egb}[1]{\ensuremath{e^{ig\,b \,#1}}}
\newcommand{\ebgm}{\ensuremath{e^{-igb}}}
\newcommand{\ebkp}[1]{\ensuremath{e^{ib^\dag \overline{\beta}(t)\,#1}}}
\newcommand{\ebkm}[1]{\ensuremath{e^{-ib^\dag \overline{\beta}(t)\,#1}}}
\newcommand{\ebp}[1]{\ensuremath{e^{ib\beta(t)\,#1}}}
\newcommand{\ebm}[1]{\ensuremath{e^{-ib\beta(t)\,#1}}}
\newcommand{\eop}{\ensuremath{e^{i\omega_0t}}}
\newcommand{\eom}{\ensuremath{e^{-i\omega_0t}}}
\newcommand{\gib}{\ensuremath{e^{-\frac{\hbar\omega_0}\theta}}}

\author{A.~M.~Sinev}
\title{Loss of the sensitivity of a gravitational wave detector due to temperature}
\date{}
\maketitle

\begin{abstract}
Using an explicit solution for the Schr\"odinger equation
describing a model of the gravitational wave detector
(LIGO-project), we prove that the SQL (standard quantum limit) for
dimensionless amplitude for gravitational perturbations of the
metric exceeds $10^{-19}$ at temperature 100K.
\end{abstract}

\noindent The first estimate of the sensitivity of a gravitational
wave detector was obtained by V.~B.~Braginsky \cite{brag}. Our
estimate of the SQL bases on the model proposed in \cite{bose,
brif_mann}. Actually, we use the Hamiltonian $\widehat H$
\cite{brif_mann}, and we find that the unitary group generated by
$\widehat H$ slightly differs from \cite{brif_mann} (see
\cite{our, vestnik}). The main observation of this paper is that
the temperature of the initial state of the oscillator decreases
the sensitivity of the measuring device.

\medskip
\noindent \textbf{1.} Let the space of the product system be
$l_2\otimes l_2$, where the first Hilbert space is the state space
of the electromagnetic field and the second factor corresponds to
the state space of the quantum oscillator. On this product space,
consider the Hamiltonian:
\begin{equation}
H=\frac{1}{\hbar}(H_0+H_{int})=\omega\,a^\dagger a\otimes I +
I\otimes \omega_0 \, b^\dagger b - (\kappa \, a^\dagger
a+f(t))\otimes (b+b^\dagger ), \label{ham}
\end{equation}
where $f(t)=\frac{F(t)}{\sqrt{2m\omega_0 \hbar}}$, $F(t)$ is a
classical force generated by a gravitational wave,
$\kappa=\frac{\omega}{L} \sqrt{\frac{2\hbar}{m\omega_0}}$ is a
coupling constant, $\omega$, $L$, $\omega_0$, $m$ are the optical
frequency, the length of the cavity, the eigenfrequency and the
mass of the oscillator correspondingly.

In paper \cite{our}, we proved that an explicit solution of the
Schr\"odinger evolution equation
$$
\frac{d}{dt}U_t=iHU_t,\quad U\bigl|_{t=0}=I\otimes I \label{probl}
$$
with the Hamiltonian (\ref{ham}) reads as follows:
\begin{equation}
U_t=e^{iH_0t}e^{-C(t)} e^{-ib^\dagger \overline\beta(t)}
e^{-ib\beta (t)}, \label{ev}
\end{equation}
where
\begin{gather*}
\beta (t)=\int_0^t\,d\tau \,(\kappa\, a^\dagger a+f(\tau))\,
e^{i\omega_0
\tau},\\
C(t)= \int_0^t\,d \tau \, (\kappa\,a^\dagger
a+f(\tau))\,e^{i\omega_0 \tau}\overline\beta (\tau).
\end{gather*}

As an observable, we choose a quantized output of the classical
two-beam interference
$$
A(t)=\frac{I_N}{2i}\biggl(e^{i\frac{\omega  x_t}{c}}-
e^{-i\frac{\omega  x_t }{c}} \biggr),
$$
where $x_t$ is the displacement of the oscillator caused by the
gravitational force:
\begin{gather}
\hat A= \frac{\hat
I_N}{2i}\biggl(e^{ig(b^\dagger+b)}-e^{-ig(b^\dagger+b)}\biggr) =
\frac{\hat I_N\, e^{-\frac{ g ^2}{2}}}{2i}
 \biggl(e^{i g  b^\dagger}e^{i g  b}-
e^{-i g  b^\dagger}e^{-i g  b} \biggr), \label{obs}\\
\hat A^2=(\hat I_N)^2\biggl\{\frac{1}{2}-\frac{e^{-2g^2}}{4}
\biggl(e^{2i g b^\dagger}e^{2i g  b}+ e^{-2i g  b^\dagger}e^{-2i g
b} \biggr)\biggr\} \notag,
\end{gather}
$ g=\frac{\omega}{c} \sqrt{\frac{\hbar }{2m\omega_0}}$, and $\hat
I_N$ is the operator corresponding to the output laser power.

If the oscillator is prepared in the ground state $\psi_0 =
\{1,0,0\ldots\}$, for the exponentially distributed number of
photons in the laser beam $\psi(z)= \{1, z / \sqrt{1!},z^2 /
\sqrt{2!}\ldots\}$, $\bar z\,z = N$, we obtain following
expressions for the mean value of the output signal (\ref{obs})
and its dispersion \cite{our}:
\begin{gather}
I(t)= \Tr_{l_2\otimes l_2} \{ \hat A \,U_t^* \psi_0 \otimes
\psi(z)\,U_t\} = I_N\,e^{-\frac{g^2-N\theta^2_l}{2}}\sin(\theta_g
+N\theta_l),\label{out}
\\
 \langle \hat A^2 \rangle = \frac{I_N^2}{2}\left(1
-e^{-2g^2-2N\theta_l^2}\cos(2\theta_g+2N\theta_l)\right), \notag
\\
D(t)=\langle \hat A^2 \rangle-I^2(t)= \notag
\\
= \frac{I_N^2}{2}\left( 1-e^{-g^2-N\theta_l^2} \right) \left(
1+e^{-g^2-N\theta_l^2}\cos(2\theta_g+2N\theta_l)
\right),\label{disp}
\end{gather}
where
\begin{equation}
\theta_g = \frac{\omega}{\omega_0} \int_0^t\frac{F_g(\tau)}{
mc}\sin\omega_0(t-\tau)\,d\tau,\quad
F_g(\tau)=Lmh_0\omega_g^2\cos(\omega_g\tau) \label{force}
\end{equation}
corresponds to the gravitational source, $\omega_g$ is the
frequency of the gravitational wave and $h_0$ is the dimensionless
amplitude of gravitational perturbations of the metric tensor, and
\begin{equation}
\theta_l=2 \kappa  g\,(1-\cos(\omega_0t)) \label{pressure}
\end{equation}
describes the light pressure of the laser beam on the oscillator.

In order to estimate the sensitivity of the detector, we use the
following condition on the mean value of the output signal
(\ref{out}) and its dispersion (\ref{disp}):
\begin{equation}
I(t)>\sqrt{D(t)}. \label{cond}
\end{equation}
In resonant case $\omega_0 = \omega_g$, this inequality gives the
standard estimate of the SQL \cite{brag} corresponding to the
ground state of the oscillator (for detailed discussion see
\cite{our}):
\begin{equation}
h_0 >
h_{SQL}=\frac{1}{Lt\omega_0}\sqrt{\frac{\hbar}{m\omega_0}}=5\cdot10^{-24}.
\label{sql}
\end{equation}
All numerical values are obtained by using realistic parameters of
LIGO II project:
\begin{equation}
\omega_g=30\, {\rm sec}^{-1}, \quad L=4\cdot 10^3\, {\rm m}, \quad
m=10\, {\rm kg}, \quad \omega=1.8\cdot 10^{15}\,{\rm sec}^{-1}.
\label{param}
\end{equation}
Estimate (\ref{sql}) coincides with the famous V. B. Braginsky
formula \cite{brag} for SQL. This solvable model allows us to
obtain the SQL for a thermal equilibrium initial state of the
oscillator.

\medskip
\noindent \textbf{2.} Suppose that the oscillator is at thermal
equilibrium with the environment at temperature $\theta = k\,T$,
where $k$ is Boltzmann constant and $T$ is Kelvin temperature.
Then the initial state of the oscillator is described by the
density matrix with Gibbs weight:
$$
\rho_0=\z \sum\limits_{n=0}^\infty e^{-\frac{\hbar \omega
n}\theta}|n\rangle\langle n|,\quad
$$
where $\displaystyle Z=\frac 1{1-e^{-\frac{\hbar\omega}\theta}}$
is the statistical sum. Let us estimate minimal detectable
perturbations of the metric, when the oscillator has non-zero
temperature before measurements.

For observable (\ref{obs}) under evolution (\ref{ev}), we obtain
the following partially averaged output signal
\begin{gather}
\hat I(t)=\hat I_N  \mathrm{Tr}_{osc}\{ e^{ig\, b^\dagger }
e^{ig\, b}U_t^* \rho _0U_t\}= \notag\\
 =\z \hat I_N\,e^{-\frac{g^2}{2}} e^{i\theta_g + i\aka\theta_l}\sum_{n=0}^\infty \sum_{k=0}^n
(-1)^k\frac{n!}{(n-k)!k!}\,\frac{g^{2k}}{k!}e^{-\frac{n\,\hbar\omega_0}\theta},
\label{sum}
\end{gather}
where $\theta_g \text{ and } \theta_l$ are given by (\ref{force})
and (\ref{pressure}). The sum in (\ref{sum}) can be calculated
explicitly. The Laguerre polynomial equals
\[
L_n(z) = L_n^{(0)}(z) = 1 - \binom n1\frac z{1!} + \binom n 2\frac
{z^2}{2!} - \ldots + (-1)^n \binom n n \frac {z^n}{n!}
\]
and their generating function is given by
$$
\frac{e^{-\frac{zt}{1-t}}}{1-t}=\sum_{n=0}^\infty L_n(z)t^n \qquad
(|t|<1).
$$
In our case, $t = \gib < 1$ and for $\langle\hat
A\rangle_{osc}=\im \hat I(t)$, we obtain
\begin{equation}
\langle\hat A\rangle=\alpha \, \hat I_N e^{-\frac{g^2}2}\sin(
\theta_g + \aka\theta_l),
\end{equation}
where
\begin{equation} \label{alpha}
\alpha=\exp \left \{-\frac{g^2}{e^{\hbar\omega/ \theta}-1}
\right\}.
\end{equation}
Considering the exponential distribution of photons in the laser
beam at large $N \sim 10^{17}$ and at small $|\theta_l|
=1.2\cdot10^{-19}$, the mean value and the dispersion of the
signal reads as follows:
\begin{gather}
I=\alpha \, I_Ne^{\frac {g^2}2}\sin(\theta_g + N\theta_l), \label{mean}\\
\langle \hat A^2 \rangle = \frac{I_N^2}{2}\left(1
-\alpha^4e^{-2g^2-2N\theta_l^2}\cos(2\theta_g+2N\theta_l)\right) \notag \\
D=\langle\hat A^2\rangle-I^2. \label{dispt}
\end{gather}
For temperatures from $10^{-9}$K to $10^{10}$K, the coefficient
$\alpha$ (\ref{alpha}) equals approximately $\alpha \approx
1-\frac {g^2\theta}{\hbar\omega_0}$. The condition for detection
of a gravitational wave~(\ref{cond}) $I(t)>\sqrt{D(t)}$ can be
estimated by linearizing expressions (\ref{mean}) and
(\ref{dispt}) since $g \ll 1$
$$
|\theta_g + N\theta_l|>\sqrt{g^2+g^2\frac{\theta}{\hbar\omega_0} +
N\theta_l^2}.
$$
If we compare the signal from the gravitational wave $\theta_g$
with its dispersion in semi-classical case, when number of photons
does not fluctuate (as was done in \cite{our}), we obtain the SQL
for perturbations of the metric (see (\ref{sql}))
\begin{equation}
h_{SQL}(T) = h_{SQL}\sqrt{1 + \frac{kT}{\hbar\omega_0}} \sim
10^{-19} \label{sqlt}
\end{equation}
at temperature $T \sim 100$K and for realistic values of the
parameters of LIGO project (\ref{param}). It is clear that thermal
noises have a great significance for the sensitivity of a
measuring device. The corresponding SQL (\ref{sqlt}) is much
higher (about $10^5$) than the SQL in vacuum (\ref{sql}).

\end{document}